\documentclass[floatfix,twocolumn,showpacs,amsmath,amssymb,aps,prb]{revtex4-1}
\usepackage{graphicx}% Include figure files
\usepackage{dcolumn}% Align table columns on decimal point
\usepackage{bm}% bold math

\newcommand{\URS}{\textrm{URu}_{1-x}\textrm{Rh}_{x}\textrm{Si}_2}

\begin{document}

\title{Local suppression of the hidden order phase by impurities in URu$_2$Si$_2$}
\author{Maria E. Pezzoli,$^{1}$ Matthias J. Graf,$^{2}$, Kristjan Haule$^{1}$, Gabriel Kotliar$^{1}$ and Alexander V. Balatsky $^{2,3}$}
\affiliation{
$^{1}$Serin Physics Laboratory, Rutgers University, Piscataway, NJ 08854, USA. \\
$^{2}$Theoretical Division, Los Alamos National Laboratory, Los Alamos, NM 87545, USA. \\
$^{3}$Center for Integrated Nanotechnology, Los Alamos National Laboratory, Los Alamos, NM 87545, USA. }

\begin{abstract}
We consider  the effects of impurities on the enigmatic hidden order (HO) state of the heavy-fermion material $\textrm{URu}_2\textrm{Si}_2$. In particular, we focus on local  effects of Rh
impurities as a tool to probe  the suppression of the HO state. To study local properties we introduce a lattice free energy, where the {\em time invariant} HO
order parameter $\psi$ and local antiferromagnetic (AFM) order parameter $M$ are
 {\em competing orders}.
Near each Rh atom the HO order parameter is suppressed, creating a hole in which local AFM order emerges
as a result of competition. These local holes are created in the fabric
of the HO state like in a Swiss cheese and ``filled'' with droplets of AFM order.
We compare our analysis with recent NMR results on
U(Rh$_x$Ru$_{1-x}$)$_2$Si$_2$ and find  good  agreement with the data.
\end{abstract}

\maketitle
\date{\today}

\section{Introduction}

The physics of heavy-fermion materials is   fascinating
and  extremely challenging due to a variety of exotic phenomena that can be observed, \emph{e.g.},
the Kondo effect, the heavy mass renormalization, the
onset of novel magnetism or of unconventional superconductivity.
The interplay between these phenomena makes a detailed understanding of the ground state   complicated.
Here we focus on  $\textrm{URu}_2\textrm{Si}_2$, the heavy-fermion material that exhibits a ``hidden order''
(HO) phase below $T_{\textrm{ho}} = 17.7$K \cite{palstra}.
The specific heat of this material displays the typical jump of a second order
phase transition at  $T_{\textrm{ho}}$, however
the  precise nature of the HO remains a subject of intensive debate.
Far above the HO transition the magnetic susceptibility has a maximum around $T \sim 50$K \cite{palstra}.
The measured magnetic moment reported by neutron scattering, if there is any, is too small ($\sim 0.03 \, \mu_B/U$)
to explain the large entropy loss at  $T_{\textrm{ho}}$ within a localized AFM scenario,
which led to the concept of the small moment antiferromagnetism\cite{broholm}.
Early $\mu$SR (muon spin relaxation) measurements reported magnetic moments as small as $\sim 10^{-3} \, \mu_B/U$\cite{MacLaughlin}.
However, later $\mu$SR and nuclear magnetic resonance (NMR) measurements on pure $\textrm{URu}_2\textrm{Si}_2$ 
revealed an inhomogeneous coexistence between the HO and AFM order with a sizable magnetic moment\cite{Luke,matsuda}.
Moreover recent neutron scattering experiments
evidence that this small moment is not an intrinsic feature of the HO, but a spurious effect due to local strains induced
by  crystal defects in the sample \cite{Nicklowitz}.
Nevertheless  magnetic ordering is not completely extraneous to $\textrm{URu}_2\textrm{Si}_2$:
an antiferromagnetic phase with large moment can be stabilized by  applying pressure or strain \cite{matsuda,yokoyama,villaume,hassinger}.
Since 1985 several theories have been proposed to identify the nature of the hidden order parameter.
Recently a resurgence of interest in this material has been seen as new data and new ideas on the nature of the HO
appeared \cite{wiebe,santander,Balatsky_CDW,Coleman_np_2009,Cricchio,Gabi1,Oppeneer,Varma}.

Further  progress  in experimental techniques such as sample quality and more accurate measurements
of $\textrm{URu}_2\textrm{Si}_2$ suggest that a breakthrough in this long standing problem is at hand and may be achieved soon.
In this paper we focus on the role of impurities
as probes of the nature of the hidden order puzzle.
We address the role of deliberately placed Rh impurities  on the suppression of the HO state.
Since  few impurities are added to the sample, NMR is a particularly useful  bulk probe  sensitive to the local atomic environment to reveal
what happens to  the HO state at the impurity site. Recently the $^{29}\textrm{Si}$ NMR spectrum
has been reported in U(Ru$_{1-x}$Rh$_{x}$)$_2$Si$_2$ as a function
of temperature and $\textrm{Rh}$ concentration \cite{Curro}.
The experiment showed local suppression  of the HO state and the emergence of satellite NMR peaks,
indicating the onset of {\em local} antiferromagnetic droplets near each Rh impurity.
These experiments were interpreted in a  Ginzburg-Landau framework, where
antiferromagnetism and hidden order are coupled through gradient terms. In this scenario
 the antiferromagnetism  is not a competing order
parameter but rather a parasitic effect induced by spatial
inhomogeneities in the hidden order parameter \cite{Curro}.

Here  we turn to a more microscopic description of the  effects of impurities
at the atomic length scale by using  a {\it lattice
free energy}, where each lattice site corresponds to a uranium atom. 
We thus extend earlier work, using a lattice free energy with
parameters describing the phase diagram of
URu$_2$Si$_2$ \cite{GabiHaule} in presence of pressure and strain,
in order to address a spatially inhomogeneous setting.
The approach is not tied however to a specific microscopic origin
of the hidden order: the form of the lattice free energy is general
and we choose a particular set of parameters values since it has been proven 
to be consistent with experiment.
In this model droplets emerge around the impurities  as a result of the
competition between the HO and AFM order, enhanced by the
coupling mechanism suggested in Ref.~\onlinecite{Curro}.

The paper is organized as follows: in Sec.~II we introduce the
lattice free energy that we minimize in order to determine the
phase diagram of U(Ru$_{1-x}$Rh$_{x}$)$_2$Si$_2$; in Sec.~III we
show our results and compare them with recent experimental
NMR data \cite{Curro}; in Sec.~IV we draw our conclusions. Finally in
appendix A we discuss how the lattice free energy can be derived
from a microscopic Hamiltonian and in appendix B we derive the
coarse-grained Ginzburg-Landau free energy from our lattice free energy model
to make contact with earlier work \cite{Curro}.

\section{Model}
 We write the free energy  in terms of two order parameters: the HO
parameter $\psi_i$  and the  AFM
order parameter $M_i$. $M_i$ is the $z$ component of the magnetic
moment, since it is observed experimentally that
$\textrm{URu}_2\textrm{Si}_2$ orders magnetically along the $z$
direction \cite{broholm,broholm_prb,matsuda}. The free energy
contains three terms: $ F =F_{\psi} +F_{M} +F_c $, where $F_c$ is
the coupling term between $\psi_i$ and $M_i$ .
 Assuming that the hidden order preserves  time-reversal symmetry,
the simplest form of $F_c$  is  $F_c = g_1 \sum_i \psi_i^2 M_i^2 \phantom{a}$ \cite{premi}.
Therefore we can write the lattice free energy  as
\begin{equation}
\label{eq:free}
\begin{aligned}
F & =  a_{\psi} \sum_i \psi_i^2 + b_{\psi} \sum_i \psi_i^4 + \frac{1}{2}   \sum_{ij} J^{\psi}_{ij}  \psi_i \psi_j \\
 &\quad + a_{M} \sum_i M_i^2 + b_{M} \sum_i M_i^4 + \frac{1}{2}  \sum_{ij} J^M_{ij}  M_i M_j \\
 &\quad + g_1 \sum_i \psi_i^2 M_i^2 \;, \\
\end{aligned}
\end{equation}
where $\psi_i$ and $M_i $ are defined for each site of a three dimensional lattice.
% Below we use the size 40 \times 40 \times 40.

The form of this lattice free energy  is  general and can accommodate different
scenarios for the hidden order  in URu$_2$Si$_2$. As discussed in
the appendices, the information about the underlying microscopic
theory is contained  in the values of the lattice free energy
parameters.
A similar phenomenological free energy was proposed for a toy model describing the competing AFM 
and hexadecapolar order emerging from crystal field splitting within the unit cell of URu$_2$Si$_2$ \cite{GabiHaule}. In
that case,  the  parameters were naturally expressed in terms of
an effective crystal field splitting $\Delta$ at each uranium site:
$a_{\psi}=a_{M}=a=\frac{\Delta}{2} \coth(\frac{\Delta}{2 T}) $,
$b_{\psi}=b_{M}=b=\frac{\Delta}{2}[\sinh(\frac{\Delta}{T}) -
\frac{\Delta}{T}]\frac{\cosh^2(\Delta/2T)}{\sinh^4(\Delta/2T)} $
and $g_1 = 2b$.  The effective exchange constants  $J^{\psi}$ and $J^{M}$
were  determined in such a way to
reproduce the experimentally observed critical temperatures,
\emph{i.e.}, the hidden order transition temperature at zero
doping $T_{\textrm{ho}}=17.7$K and the Neel temperature
$T_\textrm{N}=15.7$K, that is $J^{\psi} = \Delta /
\tanh(\Delta/2 T_{\textrm{ho}}) $ and $J^{M} = \Delta /
\tanh(\Delta/2 T_N) $ 
\footnote{
 Notice that the coefficients $J^{\psi}$ and $J^{M}$ contribute both to the massive and gradient
term in the continuum formulation. In fact in order to derive the continuous free energy, we substitute $J_{ij}=J(\vec{R})$
with the Fourier transform $J(\vec{k})= \sum_{\vec{R}} e^{i \vec{k}\cdot \vec{R}} J(\vec{R}) $ (see Appendix B).
The zero order term of $J(\vec{k})$ contributes to the massive term, while the quadratic term in $k$ to the gradient term}.
This parametrization of the lattice free energy,
using the measured elastic constants, was shown to provide  an excellent
description of  the phase diagram of URu$_2$Si$_2$ under applied
magnetic field, pressure and strain .
Here
$\Delta= 35 $K is the effective  crystal field splitting between
the two lowest lying states of the  U atom $5f$ electrons of
$\textrm{URu}_2\textrm{Si}_2 $ in the paramagnetic phase
\cite{Gabi1}.
In this work we choose to adopt the same
parametrization. We stress however that the form of this  free
energy is much more general and describes a situation where
$\psi_i$ is any order parameter that does not  break time-reversal
symmetry; for example alternative order parameters are a charge
density at \emph{incommensurate} momenta  or a hybridization wave
as proposed in Refs.~\onlinecite{Balatsky_CDW} and~\onlinecite{dubi}.

To incorporate the role of  impurities  we consider two effects. The  first is a mean-field effect in which
we regard the coefficients $a$ and $b$ to be disorder dependent.
In an itinerant picture the presence of disorder creates a random potential acting
on the electrons:  the impurities
act  as scattering centers  which reduce the excitonic pairing in the particle hole channel.
In the model of Ref. \onlinecite{GabiHaule}, the impurity induced strain
increases  the crystal field parameter $\Delta$ and therefore it suppresses both the antiferromagnetism and the hidden
order stabilizing the paramagnetic phase.
The  coefficients $a$ and $b$  thus acquire a linear
(at leading order) dependence on doping $x$ by
 imposing $\Delta = \Delta_0 + x \Delta_1 $.
However we keep the definition of $J^{\psi}$ and $J^{M}$
to be disorder independent, \emph{i.e.} 
 $J^{\psi} = \Delta_0 /
\tanh(\Delta_0/2 T_{\textrm{ho}}) $ and $J^{M} = \Delta_0 /
\tanh(\Delta_0/2 T_N) $.
\begin{figure}[!h]
\begin{center}
\includegraphics[scale=0.35]{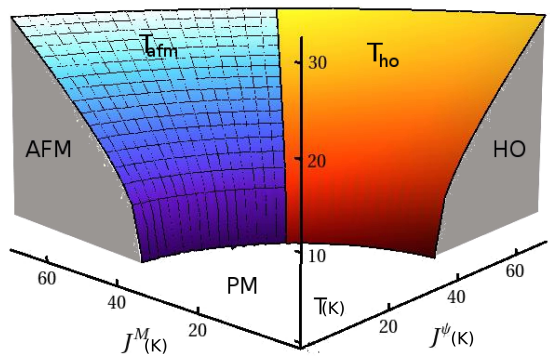}
\caption{ \label{fig:JpsiJM}(Color online) Phase Diagram of the lattice free energy described in the text 
in the  $J^{\psi}$, $J^M$, $T$ space at doping $x=0$.}
\end{center}
\end{figure}
With this parametrization there is a  duality between hidden order $\psi$  and large antiferromagnetic moment $M$.
When $J^\psi > J^M$ the hidden order phase is stabilized at low temperature; if $J^M > J^\psi$ the large moment phase is stabilized.
In Fig.~\ref{fig:JpsiJM} we show the phase diagram of the homogeneous lattice free energy of Eq.~\ref{eq:free} 
with the choice of parameters discussed above, in the $(J^{\psi}, J^M, T) $
parameter space.
The introduction of impurities, through the disorder dependence of the mean-field coefficients
$a$ and $b$, suppresses equally both HO and AFM order parameters.

Disorder also has  a very local effect: the coupling of the U which is closest to a Rh site has a different
exchange interaction with its neighbors. 
Close to Rh impurities HO is suppressed and U atoms acquire a finite moment.
 This is explicitly described by a  locally renormalized exchange term
in the hole of the swiss cheese fabric
\begin{equation}
\label{eq:freedis}
\begin{aligned}
F_{\textrm{r}_i} & =  - \frac{1}{2} [J^{M}- \tilde{J}^{M}
]\sum_{\vec{d}=\pm 1} M_{\textrm{r}_i} M_{\textrm{r}_i+\vec{d}}\
\end{aligned}
\end{equation}
where $\tilde{J}^{M} > J^{\psi}$ and by imposing $\psi(r_i) =0$.
As a result, away from the impurities $J^M < J^\psi$ and the hidden order
is stable, but in the immediate vicinity of the impurity antiferromagnetism
is stabilized over hidden order.
In order to study the local suppression of the hidden order by impurities, we
consider the lattice model described by  the free energy
\begin{equation}
\begin{aligned}
\label{eq:4}
F_{\textrm{impurities}} =  F + \sum_i F_{\textrm{r}_i} \; .
\end{aligned}
\end{equation}

To introduce a minimum number of parameters we limit  the  range of the interaction to  only the first neighbors of
the affected uranium site.
Given these definitions, the free parameters of the model are the magnetic coupling $\tilde{J}^M$ at the impurity
sites and  $\Delta_1$.
Due to the duality of the model,
as  magnetic droplets can be stabilized
in a hidden order background,  with the same mechanism 
droplets of ``hidden order'' could be stabilized within the large
moment phase by another type of impurity,  which would exchange
the role of $J^M$ and ${\tilde J}^M$. It would be interesting to
see if it is possible to realize this dual scenario
experimentally. The existence of localized regions of the HO phase in the AFM
phase at higher pressure could be observed in
NMR experiments and other local probes. The counterpart of the previous $\URS$ experiment would
require to measure $\textrm{URu}_2\textrm{Si}_2$  under pressure  to stabilize AFM but doped with
 suitable impurities to induce  a local expansion  in the ab plane.

In the numerical simulations  we will assume that the disorder is
dilute enough to  consider the solution of the mean-field
equations around a  single impurity at $r_0$.  In this work we take a simplified cubic  lattice
instead of the 
tetragonal lattice of URu$_2$Si$_2$, and we do not consider the problem
of how the order parameter on the U atoms is transferred to
the nuclear sites of the Si where the NMR is performed. Our goal
in this paper is to   explore the physics introduced by
inhomogeneities  using a lattice free energy framework and see
how the NMR experiments constrain the symmetry and the parameters in this theory.
We test whether the parametrization of the lattice free energy
that was used to successfully describe the phase diagram of
URu$_2$Si$_2$  under pressure, stress and applied magnetic field  can
also account qualitatively for the puzzling NMR measurements when
impurities are introduced in the sample.

Once we define the lattice free energy, we determine the value of $\psi_i$ and $M_i$ that minimize the
free energy, \emph{i.e.} the solutions to the equations
\begin{equation}
\begin{split}
 \frac{\delta F_{\textrm{impurities}}}{\delta \psi_i} &  = 0 \\
 \frac{\delta F_{\textrm{impurities}}}{\delta M_i} &  = 0 \,. \\
\end{split}
\end{equation}
Notice that, away from the impurity, the solutions are $M_i^2=0$ and $\psi_i^2=\frac{J^{\psi} -2a(T,x)  }{4b} $
for each site $i$ (at large  enough distance from the impurity the lattice translational invariance is  restored).

\section{Results}

We first  compute the HO critical temperature as a function of doping $x$.
Since we assume  dilute doping, we find
\begin{equation}
\label{eq:2}
 T_{\textrm{ho}}(x) = \frac{\Delta_0 + x \Delta_1}{2\, \textrm{artanh} (\frac{\Delta_0 + x \Delta_1}{J^{\psi}}) }\,.
\end{equation}
At the critical doping $x_c$ the HO parameter vanishes and $T_{\textrm{ho}}(x_c) =0 $;
 it follows that $x_c$ is given by $\Delta_0 +  x_c \Delta_1= J^{\psi} $.
\begin{figure}[!h]
\begin{center}
\includegraphics[scale=0.65]{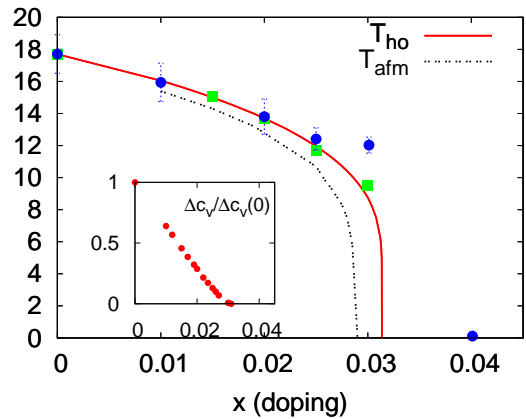}
\caption{ \label{fig:thermo}(Color online)  Computed critical temperature $T_{\textrm{ho}}$ for the order parameter $\psi$ as a
function of doping. Squares and circles are experimental values taken respectively from Ref.~\onlinecite{Curro} and
Ref.~\onlinecite{Amitsuka}. The dashed line corresponds to the temperature $T_{\textrm{afm}}$ at which the magnetic signal
disappears according to our calculation. Inset: computed jump of the specific heat
$\Delta \mathrm{c}_\mathrm{v}/\Delta \mathrm{c}_\mathrm{v}(0)$ at the phase transition as a function of doping. $\Delta \mathrm{c}_\mathrm{v}(0)$ 
is the value of the jump at zero doping.}
\end{center}
\end{figure}
In  Fig.~\ref{fig:thermo} we compare the theoretical curve
(\ref{eq:2}) for $\Delta_1 = 358.6$K with experimental data
\cite{Curro,Amitsuka}.
$J^\psi$ and $\Delta_0$ have been defined above for the uniform case and are equal to $J^\psi=46.24$K and
$\Delta_0 = 35$K.
 We observe that the transition
temperature, up to leading terms, has in our model the usual linear dependence
on  $x$ similar to impurity-averaged theories. 
In the inset of
Fig.~\ref{fig:thermo} we report the computed  jump in the
specific heat $\Delta \mathrm{c}_\mathrm{v} = \frac{-1}{V}\frac{\partial^2
F}{\partial T^2}$. Since the free parameter $\Delta_1$ was determined
by the critical  temperature the agreement with the experimental
data is very satisfactory \cite{Curro}. In Fig.~\ref{fig:thermo} we plot
also $T_{afm}$ as a function of doping. $T_{afm}$ is defined as
the temperature at which the magnetization becomes smaller than the minimum
observed magnetic moment $\mu_{0,\textrm{min}} \approx 0.03 \; \mu_B$.

\begin{figure}[!h]
\begin{center}
\includegraphics[scale=0.65]{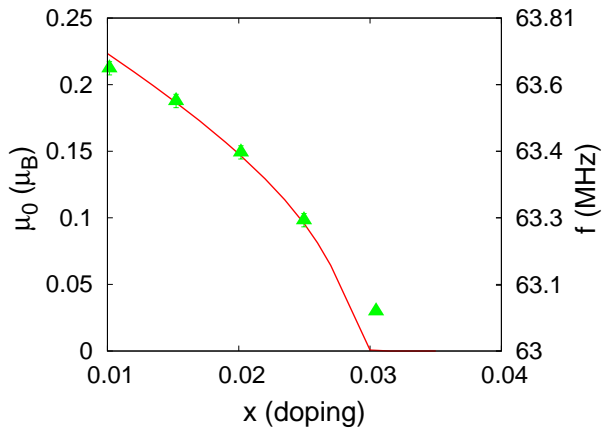}
\caption{ \label{fig:freq} (Color online) Magnetic moment induced by the Rh impurities, in units of $\mu_B$ (left  axis) and in units
of the corresponding NMR frequency (right  axis) as a function of $x$ at $T=4$ K.
The frequencies of the two satellite peaks are $ f_{\pm}= \gamma (H_0 \pm A \mu_0)(1+ K)  $,
where $\gamma=8.46 \,\mathrm{MHz}/\mathrm{T} $ is the gyromagnetic ratio of $^{29}\textrm{Si}$, $K=0.065$ is the
Knight shift, $H_0= 7 \,\mathrm{T}$ is the external field, $A=0.36\, \mathrm{T}/\mu_B$ is the hyperfine coupling
and $\mu_0 $ is the ordered spin moment of U atoms.
 The red line is the result of our model, while full triangles
are experimental points taken from Ref.~\onlinecite{Curro}.
The computed magnetic moment is  the value of the magnetization at the impurity site.}
\end{center}
\end{figure}
In Fig.~\ref{fig:freq} we report the NMR frequencies $f$ as a function of Rh concentration $x$.
The NMR frequency $f$ is proportional to the spin moment $\mu_0$ at each U site, in particular
$ f_{\pm}= \gamma (H_0 \pm A \mu_0)(1+K)  $,
where $\gamma=8.46 \,\mathrm{MHz}/\mathrm{T} $ is the gyromagnetic ratio of $^{29}\textrm{Si}$, $K=0.065$ is the
Knight shift, $H_0= 7 \,\mathrm{T}$ is the external field, $A=0.36\, \mathrm{T}/\mu_B$ is the hyperfine coupling.
In our model we identify the magnetization at the impurity site $M_{r_0}$ with $\mu_0$.
The theoretical curve in Fig.~\ref{fig:freq} is obtained with $\tilde{J}^M= 33.83 $K.
The emergence of a  magnetic moment in $\textrm{URu}_2\textrm{Si}_2$ doped
with Rh has been observed also in neutron scattering experiments \cite{Amitsuka}.
Taking into account that neutrons measure the magnetization averaged over volume, the agreement
between the two measures is good.
From the NMR spectrum we obtain another valuable information: the area under the satellite peaks is proportional
to the fraction of antiferromagnetic sites, \emph{i.e.} the ratio between sites with a finite magnetization and the total
number of sites. The measured antiferromagnetic fraction has a non-monotonic behavior as a function of the
Rh concentration $x$:  first it increases linearly with the
number of impurities, then it reaches a maximum at $x=0.025$ and finally it decreases becoming zero after the critical
concentration has been reached \cite{Curro}. 
The neutron scattering result is consistent with this observation: 
the intensity of the magnetic Bragg peaks
has a non-monotonic behavior as a function of $x \;$ \cite{matsuda}. 
Our model offers a simple explanation of this non-monotonic behavior.
\begin{figure}[!h]
\begin{center}
\includegraphics[scale=1.5]{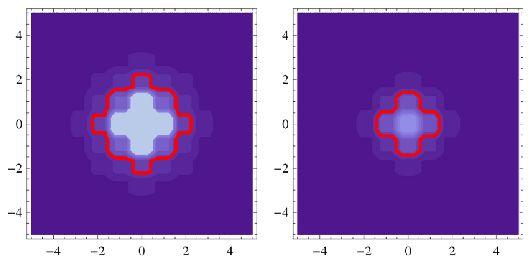}
\includegraphics[scale=0.8]{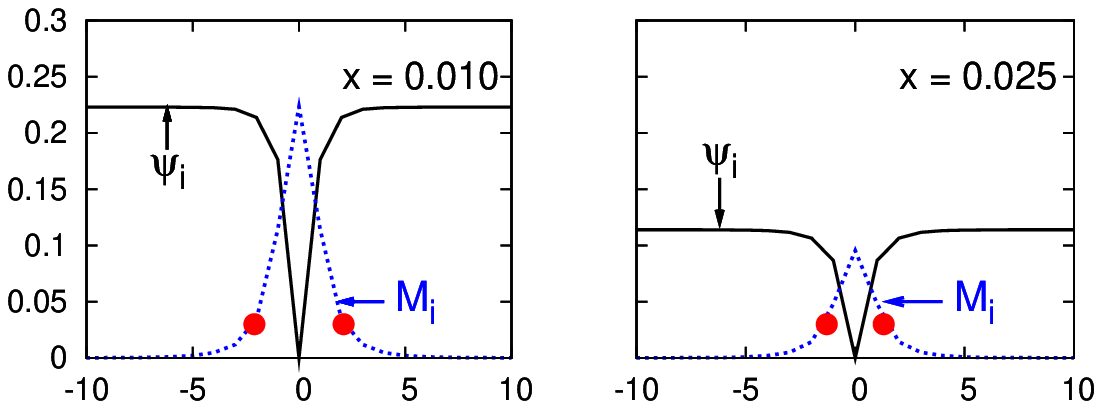}
\caption{ \label{fig:droplet} (Color online) Upper panels: magnetization density in the plane $z=0$ at $T=4$K for
Rh concentration $x=0.01$ (left) and $x=0.025 $ (right).
We display the magnetization in the interval $[0.002,0.22]$: brighter colors correspond to higher values.
 The red contour
corresponds to the magnetic droplet boundary as defined in the text.
Lower panels: profile of the magnetization $M_i$ (blue dashed line) and of the HO order parameter $\psi_i$ (black solid line)
along the direction (0,y,0) of the lattice. Red dots correspond to the intersection of the magnetization with droplet boundaries. }
\end{center}
\end{figure}
In Fig. ~\ref{fig:droplet}  a real-space representation of the magnetization $M_i$ at each site
is reported for impurity concentrations $x=0.01$ and $x=0.025$.
The sites around the impurity develop a finite moment.
 The moment at each site decreases  as  the distance  from the impurity
increases. The magnetization is strongly suppressed and $\psi_i$
recovers the bulk solution value within few lattice sites, see
profile picture in Fig.~\ref{fig:droplet}. In the following we
will refer to the magnetic sites around the impurity with the
term ``droplet''. We define the droplet boundary in such a way
that  the magnetization of every site inside the droplet is large
enough to be observed experimentally. We consider that the
minimum observed magnetic moment is equal to
$\mu_{0,\textrm{min}}=0.03 \; \mu_B$. We can see in Fig.~\ref{fig:droplet}
that the size of the droplet is  affected by
disorder. The two competing effects of disorder are evident: on
the one hand the number of magnetic droplets increases with the
number of impurities, on the other hand in our model the size of each droplet
decreases with increasing disorder. 
This leads to the observed
non-monotonicity in the experiments.
Notice that, in a model where HO and AFM order are coupled through 
a term $M^2 |\vec{\nabla} \psi|^2 $, the behavior of magnetic droplets
in function of doping is similar \cite{Curro}.
 In order to put this
analysis on  more quantitative grounds, we optimize the free
energy for different values of temperature and Rh concentration
and then we compute the fraction of sites with $M_i \neq 0$ ($M_i
> \mu_{0,\textrm{min}}$).
 Since we made the assumption  that
 magnetic droplets are disjoint with average spacing $d \sim l/x^{1/3} > 360$ nm (here $l$ is the lattice constant),
 we can define the antiferromagnetic fraction in the following way
\begin{equation}
 \textrm{AFM fraction} = \frac{n_{\textrm{in}} \times n_{\textrm{imp}}}{N_{\textrm{tot}}}=  n_\textrm{in} \times x \; ,
\end{equation}
where $n_{\textrm{in}}$ is the number of sites inside the droplet, $n_{\textrm{imp}} $ is the number of impurities and $N_{\textrm{tot}}$ is the total
number of sites.
\begin{figure}[!h]
\begin{center}
\includegraphics[scale=0.75]{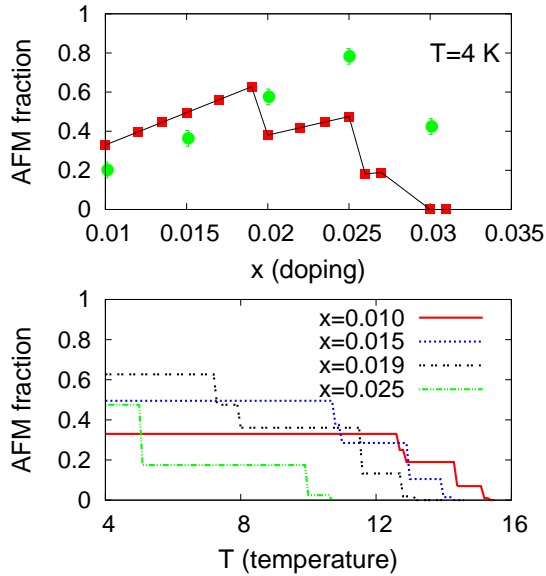}
\caption{ \label{fig:fraction}
(Color online) Upper panel: comparison between the measured (green circles) and the computed (red squares) AFM fraction, as defined
in the text, for different values of Rh concentration at temperature $T=4$K. The black line is a guide to the eye. Lower panel: computed AFM fraction as a function of temperature $T$ for different values of the Rh concentration $x$. }
\end{center}
\end{figure}
In figure \ref{fig:fraction}, upper panel, we plot the AFM
fraction as a function of doping. The curve has the
characteristic non-monotonic behavior of the
experimentally observed AFM fraction, which highlights the two competing effects of disorder: the chemical
pressure and the suppression of  order. Crucial to this observation is the duality between
$\psi$ and $M$, and hence the fact that Rh impurities suppress both order parameters.
 While the  sawtooth profile is a 
consequence of our lattice model, there is good agreement between
 the  results  of the minimization
procedure and experiment.
The sawtooth profile appears because the number of sites with
$M_i > \mu_{0,\textrm{min}} $ is a step function of $x$. In fact
lowering the temperature below $T_{N}$ the first site to be
magnetized is the impurity site $r_0$, then the 
nearest neighboring
sites, followed by the next-nearest neighbors and so on. Therefore the droplet size increases (and with
the same mechanism the droplet decreases as a function of doping)
 in steps equal to the coordination number. In the lower panel of figure
\ref{fig:fraction} we plot the AFM fraction as a function of
temperature. At low temperature the AFM fraction is a
non-monotonic function of $x$ as  discussed above; increasing the
temperature the magnetic droplet can be stabilized only for lower
doping values. We identify with $T_{\textrm{afm}}$ the
temperature in correspondence to the disappearance of the
magnetic droplet for a given Rh concentration.  We observe
that at large doping  $T_{\textrm{afm}}$  follows the
behavior of $T_{\textrm{ho}}$ as a function of $x$, see
Fig.~\ref{fig:thermo}.

\section{Conclusion}

In conclusion, we proposed an analysis within a mean-field
lattice free energy to reveal the local  competition between  HO
and AFM phase. We introduced disorder in the model as the driving
force of two competing effects: the local stabilization of
magnetization and the suppression of both the HO and AFM order by the
impurity. We recovered the main features of the phase diagram and
the non-monotonic behavior of the AFM volume observed
experimentally. Moreover, we found that the healing lengths of
$\psi_i$ and $M_i$ are on the scale of the local strains that
stabilize the magnetization.  An additional effect
present in the calculation is the stabilization  of the
phases due to inhomogeneities pointed out in Ref. \onlinecite{Curro},
which is manifest in the continuum theory supported by our
lattice model as shown in  Appendix B.
Finally our model describes a duality between HO and AFM order: as magnetic droplets
can be stabilized by impurities in the HO phase, with a similar mechanism  HO droplets 
can be formed in a magnetic phase.

 We used in this paper a
classical lattice free energy.  Since the phase transitions occur
at rather low temperatures, it would be interesting to extend our
work to include quantum effects, including effects which would be
described by time derivatives of the order parameter as well as
the effects of damping  due to particle-hole excitations. The study
of these effects as well as their derivation from microscopic
models will provide further constraints on the possible origin of
the hidden order state in URu$_2$Si$_2$.
This would allow also a more refined
modeling of the NMR line-shapes and the mechanism for the
transfer of the hyperfine fields from the uranium to the ligand site.

Acknowledgments: We are grateful to N. Curro, S. H. Baek, J.C. Davis, and
Marcel Porta for useful discussions. This work was supported by
US DOE, under BES and UCOP010 funding. GK and MP acknowledge the
support of DOE BES DE-FG02-99ER45761   and subcontract
83509-001-10 to Rutgers. KH was supported by  the ACS Petroleoum
Research Fund 48802 and Alfred P. Sloan foundation.

\appendix

\section{Mapping from microscopic Hamiltonian onto the lattice free energy}
A general and formal expression for  the lattice
free energy in terms of the HO and AFM order parameters $\psi_i$ and $M_i$ is
given by
\begin{widetext}
\begin{equation}
\begin{split}
 Z & = \int \textrm{d} \lambda_1 \int \textrm{d} \lambda_2
\int \mathcal{D}\Psi  \mathcal{D} \Psi^{\dagger}
e^{- \int \textrm{d} x \mathcal{L} [\Psi^{\dagger},\Psi] + \sum_{i} \lambda_{1,i}
\left( O_1( [\Psi^{\dagger}\Psi]_i) - \psi_i \right) + \sum_{i}  \lambda_{2,i} \left( O_2( [\Psi^{\dagger}\Psi_i]) - M_i \right) } \\
& = e^{- \beta F[\psi_i,M_i]}
\end{split}
\end{equation}
\end{widetext}
The Lagrangian $\mathcal{L}[\Psi^{\dagger},\Psi] $ is the
starting point of the calculation written in terms of creation
and annihilation operators containing all the relevant bands. The
starting point  can be  simplified  depending on the  itinerant or localized model one considers and on  the full   quantum many-body Hamiltonian. 
However, in the general case the evaluation of the free energy is more involved and  will not be attempted here.  $\psi$ and $M$ are two order parameters with $\psi$
time reversal invariant and $M$ breaking time reversal symmetry,
the index $i$ runs over the lattice positions of the U atoms.
Different proposals of the hidden order differ in the definition
of the operator $O_1$. In the proposal of
Ref.~\onlinecite{Balatsky_CDW} the hidden order is a charge
density wave with incommensurate wave vector $\vec{Q}^*$, and in
this case the condensate  is related to the Fourier transform of
$O_1( [\Psi^{\dagger}\Psi]_i)$
\begin{equation}
 O_1( [\Psi^{\dagger}\Psi])= \sum_{\vec{k},\sigma \sigma'}\Psi^{\dagger}_{\sigma}(\vec{k}-\vec{Q}^*)\Psi_{\sigma'}(\vec{k})_{\sigma'} \delta_{\sigma,\sigma'} \,.
\end{equation}
For a hexadecapolar order as in Ref.~\onlinecite{GabiHaule} the
operator $O_1$ is equal to
\begin{equation}
 O_1([\Psi^{\dagger}\Psi])= C \sum_{kk'} \sum_{\sigma \sigma'} F_{\sigma \sigma'}(\vec{k},\vec{k}') \Psi_{\sigma}^{\dagger}(\vec{k} - \vec{Q}) \Psi_{\sigma'}(\vec{k}')
\end{equation}
where $C$ is a normalization constant, $\vec{Q}$ is the commensurate
ordering vector   and $F_{\sigma \sigma'}(\vec{k},\vec{k}') $
are defined by
\begin{widetext}
\begin{equation}
\begin{split}
 F_{\uparrow \downarrow}(\vec{k},\vec{k'}) & = \frac{3 \sqrt{5}}{64 \pi} \left[ 5  \sqrt{7} (k_x - i k_y)^3 (k{'2}_x- i k'_xk'_y - k_y^{'2})k'_z
- (k_x^2 + i k_x k_y - k_y^2)k_z(i k_y' + k'_x)(-1 + 5 k_z^{'2})  \right]\gamma_{k} \gamma^*_{k'} \\
 F_{\downarrow \uparrow}(\vec{k},\vec{k'}) & = \frac{3 \sqrt{5}}{64 \pi} \left[ 5 \sqrt{7} (k_x + i k_y)^3  (k^{'2}_x+ i k'_xk'_y - k_y^{'2})k'_z
                                      - (k_x^2 - i k_x k_y - k_y^2)k_z( k'_x - ik'y)(-1 + 5 k_z^{'2})  \right] \gamma_k \gamma_{k'}^{*} \\
 F_{\uparrow \uparrow}(\vec{k},\vec{k'}) & = \frac{3 \sqrt{5}}{64 \pi} \left[ 5 (k_x^2 - i k_x k_y - k_y^2)k_z(k_x^{'2} - i k'_x k'_y - k_y^{'2})
k'_z - (k_x - i k_y)^3 (k'_x - i k'_y)(-1+ 5 k^{'2}_z) \right] \gamma_k \gamma_{k'}^{*}  \\
F_{\downarrow \downarrow}(\vec{k},\vec{k'}) & = \frac{3 \sqrt{5}}{64 \pi} \left[ 5 (k_x^2 + i k_x k_y - k_y^2)k_z(k_x^{'2} + i k'_x k'_y - k_y^{'2})
k'_z - (k_x + i k_y)^3 (k'_x + i k'_y)(-1+ 5 k^{'2}_z) \right] \gamma_k \gamma_{k'}^{*} \, , \\
\end{split}
\end{equation}
\end{widetext}
with $\gamma_{k} = 4 \pi \int \textrm{d} r \;r^2 j_3(k r) R(r)$. In
the definition of $\gamma_{k}$, $R(r)$  is the radial wave function
of the $f$-electrons and $j_3(k r)$ is the spherical Bessel
function of order 3.
 The operator $O_2$ is the magnetization operator
\begin{equation}
\begin{split}
 & O_2([\Psi^{\dagger}\Psi]) = \frac{- 2 \mu_B}{Q^2} 
\sum_{\vec{k} \vec{k'},\sigma,\sigma'}  
 \int \textrm{d} \vec{r} e^{-i \vec{Q}\cdot \vec{r}} e^{-i \vec{k}\cdot \vec{r}}
\Psi_{\sigma}^{\dagger}(\vec{k}) \\
\times & \big \{ \vec{Q} \times \left[ \frac{1}{2} \vec{\sigma}_{\sigma,\sigma'} \times \vec{Q} + \delta_{\sigma,\sigma'} \vec{\nabla}\right]  
 e^{i \vec{k'}\cdot{\vec{r}}}  \big \} \\
 \times & \Psi_{\sigma'}(\vec{k}')
\end{split}
\end{equation}
where $\vec{\sigma}$ are the Pauli matrices and $\vec{Q}$ is a reciprocal lattice vector.

From a lattice free energy perspective different
\emph{microscopic} models
result in different values of the
coefficients. 
An important coefficient is the coherence length;
from the numerical simulation we estimate the coherence length
for the magnetic droplet
to be $\sim 3$ lattice constants at $T= 4$K and doping $x=0.01$.
The NMR data place important constraints on this parameter given
that more itinerant models give rise to longer coherence lengths
and more diffuse domain walls for the order parameter defined on
the lattice scale.

\section{Mapping from lattice free energies to coarse-grained 
Ginzburg Landau free energy}

The free energy of Eq.~(\ref{eq:free}) defines a model on the
\emph{lattice} describing the $5f$-U electrons in   URu$_2$Si$_2$.
In the continuum a  Ginzburg-Landau (GL) free energy can be derived from a  lattice
model by suitable coarse graining.  The  GL
description is formulated in terms of slowly varying amplitude
fields $\phi(x)$. Here we describe the coarse
graining starting from the  microscopic free energy  $ F$ used in
Ref.~\onlinecite{GabiHaule}. We  keep higher order terms in the coupling
between the hidden order parameter and the magnetization but
focus only on the form of the GL action to connect
it to the earlier work of Refs. \onlinecite{premi,Curro}.

 The starting point is the   free
energy~(\ref{eq:free}) on a cubic lattice introduced in Ref. \onlinecite{GabiHaule}
\begin{equation}
\begin{split}
\label{eq:freeterm}
 F[\psi_i,M_i,h^{\psi}_i,h^{M}_i] & = \frac{1}{2} \sum_{ij} J_{i,j}^{\psi} \psi_{i} \psi_{j} - \sum_{i} h_i^{\psi} \psi_i \\
+  \frac{1}{2} \sum_{ij}  J_{i,j}^{M} & M_{i} M_{j} - \sum_{i} h_i^{M} M_i \\
-  \frac{1}{2} T \sum_i  \log & \left(  \textrm{cosh}\left(\frac{1}{T} \sqrt{\left(\frac{\Delta}{2}\right)^2 + (h^{\psi}_i)^2 + (h^{M}_i)^2}\right)\right) \\
\end{split}
\end{equation}
written in terms of the order parameters $\psi_i$, $M_i$ and of
the molecular
 Weiss fields $h^{\psi}_i$, $h^{M}_i$, see
Ref.~\onlinecite{GabiHaule}. At the extrema of the free energy
$h^{\psi}_i$ and $\psi_i$ satisfy the relations
$h^{\psi}_i=\sum_j J_{ij}^{\psi} \psi_j $ and $\psi_i = -
\frac{h^{\psi}_i}{2} \textrm{tanh}(\lambda_i/T \lambda_i) $ with
$\lambda_i = \sqrt{(\Delta/2)^2 + (h^{\psi}_{i})^2 +
(h^{M}_{i})^2}$ ~\cite{GabiHaule}. The same equations are
satisfied by $h^{M}_i$ and $M_i$. In Eq.~(\ref{eq:freeterm}) we
write $h^{M}_i$ and $h^{\psi}_i$ in terms of $M_i$ and $\psi_i$
exploiting the above relations and expand the free energy
neglecting terms of the order of $O\left((\sum_j J^{\psi}_{ij}
\psi_{j})^2 + (\sum_j J^{M}_{ij} M_{j})^2  \right)^3$\, and
higher. The Fourier transforms of the lattice variables $\psi_i$
and $M_i$ are $\psi(\vec{k}) = \frac{1}{\sqrt{N}} \sum_i e^{i
\vec{k}\cdot \vec{R}_i} \psi_i $ and $M(\vec{k}) =
\frac{1}{\sqrt{N}} \sum_i e^{i \vec{k}\cdot \vec{R}_i} M_i $.
Since our mean-field description includes only nearest neighbors
antiferromagnetic coupling, the modes that condense are
$M(\vec{k}=\vec{Q})$ or $\psi(\vec{k}=\vec{Q}) $ with
$\vec{Q}=(\pi,\pi,\pi)$, depending on the relative size of
$J^{\psi}$ and $J^{M}$. The free energy of
Eq.~(\ref{eq:freeterm}) can be rewritten in terms of
$\psi(\vec{k})$, $M(\vec{k})$ and the coupling constants
$J^{\psi(M)}(k)$. In order to obtain the free energy in the
GL form we keep only the modes with $\vec{k}$ close
to $\vec{Q}$ \cite{Amit}. For sake of simplicity we shift the
wave vectors of the Brillouin zone by
 $\vec{Q}$ and therefore we consider only the modes close to $\vec{k}=0$ ($k < \Lambda$).
For small $\vec{k}$ values, the Fourier transform of the coupling
constant $J^{\psi(M)}(k)$  can be approximated as
\begin{equation}
 J^{\psi(M)}(k)=J-\frac{1}{2} J k^2 + O(k)^4
\end{equation}
where we scaled $J$ as $J/z$, $z$ being the coordination number.
After writing the free energy in terms of $\psi(\vec{k})$ and
$M(\vec{k})$ and keeping only the modes with $k < \Lambda$,
 we go back to a real-space representation using the transformation
\begin{eqnarray}
 \phi_1(\vec{r})& = & \frac{1}{\sqrt{V}} \sum_{k < \Lambda} e^{-i \vec{k} \cdot \vec{r}} \psi(\vec{k}) \label{eq:psicont} \\
 \phi_2(\vec{r})& = & \frac{1}{\sqrt{V}} \sum_{k < \Lambda} e^{-i \vec{k} \cdot \vec{r}} M(\vec{k})    \label{eq:Mcont} \,.
\end{eqnarray}
Indeed in Eqs.~(\ref{eq:psicont}) and~(\ref{eq:Mcont}) the sum is
on a discrete non-periodic set of wave vectors $\vec{k}$ with $k
< \Lambda$, therefore the fields $\phi_1(\vec{r})$ and
$\phi_2(\vec{r})$ are  slowly varying and continuous. In terms of
the fields  $\phi_1(\vec{r})$ and $\phi_2(\vec{r})$ the free GL
energy becomes
\begin{eqnarray*}
 F & = & \int \textrm{d} \vec{r} \sum_{\alpha=1,2} \left( \frac{1}{2} \mu_{\alpha}(T) \Big( \phi_{\alpha}(\vec{r}) \Big)^2 
+ \frac{1}{2} k_1 \left \lvert  \vec{\nabla} \phi_{\alpha}(\vec{r})  \right \lvert^2 \nonumber \right) \\
+ &  \frac{1}{4} & \; u \sum_{\alpha,\beta=1,2} \big(\phi_{\alpha}(\vec{r})\big)^2 \big(\phi_{\beta}(\vec{r})\big)^2 \nonumber \\
- & \frac{1}{4} & \; k_2 \sum_{\alpha,\beta=1,2}  \left \lvert \vec{\nabla}
\phi_{\alpha}(\vec{r}) \right \lvert^2
\left( \phi_{\beta}(\vec{r}) \right)^2  \nonumber \\
- & \frac{1}{2} & \; k_2 \sum_{\alpha,\beta=1,2} \sum_{ij} \delta_{ij} \left(  \partial_i \phi_{\alpha}(\vec{r}) \right) \left(\partial_j \phi_{\beta}(\vec{r}) \right)
\phi_{\alpha}(\vec{r}) \phi_{\beta}(\vec{r}) \,.
\end{eqnarray*}

The coefficients $k_1$ and $u$ are definite positive. To obtain the
traditional form of the free energy we restrict the temperature
dependence to the coefficient $\mu(T)$ of the quadratic term. The
 coupling coefficients are the same for
$\phi_1(\vec{r})$ and $\phi_2(\vec{r})$ since this GL free energy has been derived from a microscopic model
where hidden order and magnetization are related to each other,
however the form of the free energy is completely general, and
for the URu$_2$Si$_2$ system was first discussed in the work of Ref.
\onlinecite{premi}.  For a sufficiently repulsive quartic interaction,
it captures the competition and interplay between the HO and AFM
order: the field $\phi_2(\vec{r})$ can develop only if the hidden
order field $\phi_1(\vec{r})$ is suppressed.
The gradient coupling term $k_2  (\phi_2(\vec{r}))^2
\lvert \vec{\nabla} \phi_1(\vec{r})\vert^2$   was introduced and discussed
in detail in Ref. \onlinecite{Curro} to explain the non-monotonic behavior of the antiferromagnetic fraction
in the NMR spectrum. When  $k_2 < 0$
inhomogeneities in $\phi_1$  can  nucleate 
a parasitic second order parameter $\phi_2$ near impurities
even when $\mu_2>0$.

\bibliography{paper}

\end{document}